\documentstyle [12pt] {article}
\input epsf

\newcommand{\beq}{\begin{equation}}
\newcommand{\eeq}{\end{equation}}
\newcommand{\beqs}{\begin{eqnarray}}
\newcommand{\eeqs}{\end{eqnarray}}

\begin{document}

\def\thefootnote{\fnsymbol{footnote}}
\baselineskip 6.0mm

\begin{flushright}
\begin{tabular}{l}
ITP-SB-96-18   \\
May, 1996
\end{tabular}
\end{flushright}

\vspace{8mm}
\begin{center}
{\Large \bf Some New Results on Complex-Temperature } 

\vspace{2mm}

{\Large \bf Singularities in Potts Models on the Square Lattice}

\vspace{16mm}

\setcounter{footnote}{0}
Victor Matveev\footnote{email: vmatveev@insti.physics.sunysb.edu}
\setcounter{footnote}{6}
and Robert Shrock\footnote{email: shrock@insti.physics.sunysb.edu}

\vspace{6mm}
Institute for Theoretical Physics  \\
State University of New York       \\
Stony Brook, N. Y. 11794-3840  \\

\vspace{16mm}

{\bf Abstract}
\end{center}
    We report some new results on the complex-temperature (CT) singularities
of $q$-state Potts models on the square lattice.  We concentrate on the
problematic region $Re(a) < 0$ (where $a=e^K$) in which CT zeros of the
partition function are sensitive to finite lattice artifacts.  From analyses
of low-temperature series expansions for $3 \le q \le 8$, we establish the
existence, in this region, of complex-conjugate CT singularities at which
the magnetization and susceptibility diverge.  From calculations of zeros
of the partition function, we obtain evidence consistent with the inference
that these singularities occur at endpoints $a_e, \ a_e^*$ of arcs protruding
into the (complex-temperature extension of the) FM phase. Exponents for these
singularities are determined; e.g., for $q=3$, we find $\beta_e=-0.125(1)$,
consistent with $\beta_e=-1/8$.  By duality, these results also imply
associated arcs extending to the (CT extension of the) symmetric PM phase.
Analytic expressions are suggested for the positions of some of these
singularities; e.g., for $q=5$, our finding is consistent with the exact
value $a_e,a_e^*=2(-1 \mp i)$.  Further discussions of complex-temperature
phase diagrams are given.

\pagestyle{empty}
\newpage

\pagestyle{plain}
\pagenumbering{arabic}
\renewcommand{\thefootnote}{\arabic{footnote}}
\setcounter{footnote}{0}

\section{Introduction and Model}

   In this paper, we report some new results on complex-temperature 
singularities of the $q$-state Potts model \cite{potts, domb} on the 
square lattice.  The Potts model has been of interest both as an example
of a particular universality class for critical phenomena and as a model for
physical phenomena such as the adsorption of certain gases on 
substrates \cite{adsorp}.  However, in contrast to the 2D Ising model 
(equivalent to the $q=2$ case) the free energy of the Potts model for general 
$q$ has never been calculated in closed form, even for zero external field(s).
Some exact results have been established for the model: from a duality 
relation, the critical point separating the disordered, $Z_q$-symmetric 
high-temperature phase from the low-temperature phase with spontaneously 
broken $Z_q$ symmetry and associated nonzero ferromagnetic (FM) 
long-range order is known \cite{potts}.  The free energy and latent heat
\cite{baxter73}, and magnetization \cite{baxter82} have been calculated 
exactly by Baxter at this critical point, establishing that the model has a 
continuous, second-order transition for $q \le 4$ and a first-order 
transition for $q \ge 5$.  Baxter has also shown that although the $q=3$ model
has no phase with antiferromagnetic (AFM) long-range order at any finite 
temperature, there is an AFM critical point at $T=0$ \cite{baxter82}.  The
values of the critical exponents (for the range of $q$ where the transition is
continuous) have been determined \cite{expon}.  Subsequently, further insight
into the critical behavior was gained using the methods of conformal field 
theory \cite{cft}.  A review of work up through 1982 was given in Ref. 
\cite{wurev}. 

In general, if one knew the exact (zero-field) free energy, one would be able 
to determine the full phase diagram as a function of complex temperature. The
idea of generalizing a variable on which the free energy depends from real
physical values to complex values was pioneered by Yang and Lee \cite{yl}.  
These authors
considered the generalization of the external magnetic field to complex values
\cite{yl} and proved a celebrated theorem that the complex-field zeros of the 
Ising model partition function lie on the unit circle in the $\mu$ plane, 
where $\mu=e^{-2\beta H}$, pinching the real axis as the temperature $T$ 
decreases through the critical point $T_c$.  Complex-temperature (CT) 
singularities of Ising models, first considered in Ref. \cite{fisher}, were 
investigated both by means of CT zeros of the partition function 
\cite{katsura}-\cite{abe} and via their effects on low-temperature series 
expansions \cite{dg}.  As well as being of historical interest, these are
relevant here because of the equivalence of the (spin 1/2) Ising model and
$q=2$ Potts model. There is continuing interest in such complexifications 
because of the deeper insight which they give one into the properties of 
statistical mechanical models (for the Ising model, see, e.g., Refs. 
\cite{g75}-\cite{jge}).  From general arguments and comparisons with exact
solutions for 2D Ising models with isotropic couplings, one knows that in the 
thermodynamic limit, CT zeros merge together to form curves (including possible
line segments) across which the free energy is non-analytic.  These curves form
the complex-temperature phase boundary (CTPB) ${\cal B}$ of the model.  One can
define notions of complex-temperature extensions (CTE's) of the physical
paramagnetic (PM), ferromagnetic (FM) and (if it occurs) antiferromagnetic 
(AFM) phases. In certain cases there are other (labelled ``O'') 
complex-temperature phases which 
do not have any overlap with any physical phase.  These various CT phases are
separated by boundaries comprising ${\cal B}$.  The locus of points making up
${\cal B}$ may also contain part(s) consisting of curves (arcs) 
or line segments which protrude into and terminate in, certain phases.  

   There have been several calculations of complex-temperature zeros of the 
partition function for the Potts model on the square lattice
\cite{mr}-\cite{wuetal}. Since the early calculations for $q=3,4$, it has been 
recognized that the zeros show
one clear feature: if one uses duality-preserving boundary conditions, then 
in the $Re(a) > 0$ region (where $a=e^K$; see below for notation), these 
zeros lie on a portion of the unit circle $|x|=1$, where $x=(a-1)/\sqrt{q}$ 
\cite{mm}-\cite{chw}. In passing, we note that in 
the $q \to \infty$ limit it has been shown (assuming that the 
$q \to \infty$ limit and the thermodynamic limit commute) that the CT zeros 
lie on the unit circle $|x|=1$ \cite{wuetal,qinf}.  
However, for a given (finite) $q$, the situation in the $Re(a) < 0$ region has
proved to be much more difficult to elucidate. The zeros exhibit 
considerable scatter and, as we shall demonstrate, significant sensitivity to 
the boundary conditions used for the finite lattice calculations, even if one
requires these to preserve duality.  These facts have rendered it problematic
to try to make inferences from calculations of zeros on finite lattices about
the complex-temperature phase boundary ${\cal B}$ in the thermodynamic limit.
In one early work \cite{mr} it was conjectured that in the thermodynamic 
limit the zeros lie on the two circles $|a-1|=\sqrt{q}$ and 
$|a+1|=\sqrt{4-q}$ for $q=3$ and $q=4$ (where the second circle degenerates to
a point), but shortly
thereafter, from a calculation of zeros for the $q=3$ model on larger lattices,
it was concluded that this conjecture was false \cite{mm} and the zero
distribution did not suggest the existence of any simple algebraic expression
which would describe this distribution.  The same conclusion was reached 
 from a calculation of the zeros for the $q=4$ model \cite{m1}. 

   We have been able to make progress in the problematic $Re(a) < 0$ region by
employing a powerful method not hitherto used for this purpose, viz., to
combine analyses of low-temperature series with calculations of CT zeros of the
partition function.  We report our results here. In our series work, we have 
taken advantage of the recent calculations of quite
long low-temperature series for the free energy, magnetization, and
susceptibility of the square-lattice Potts model for $q=3$
up to $q=10$ by Briggs, Enting, and Guttmann \cite{beg}, extending earlier 
calculations (e.g., Ref. \cite{kihara}; for $q=3$ and $q=8$, see also Ref. 
\cite{creutz}.) The organization of this paper is as follows.
In section 2 we define the model and our notation and mention some of the
general exact results which are known.  In sections 3 and 4 we present our 
new results for the $q=3$ and $q=4$ Potts model.  In section 5 we mention some
similar results for $q \ge 5$ Potts models.  Concluding remarks are given 
in section 6. 

\section{Definition of Model and Exact Results} 

The (isotropic, nearest-neighbor) $q$-state Potts model on a lattice $\Lambda$
is defined, at a temperature $T$, by the partition function 
\beq
Z = \sum_{ \{ \sigma_n \} } e^{-\beta {\cal H}}
\label{zfun}
\eeq
with the Hamiltonian
\beq
{\cal H} = -J \sum_{\langle nn' \rangle} \delta_{\sigma_n \sigma_{n'}} 
- H \sum_n \delta_{0 \ \sigma_n }
\label{ham}
\eeq
where $\sigma_n=0,...,q-1$ are $Z_q$-valued variables on each site $n \in
\Lambda$; $\beta = (k_BT)^{-1}$; and $\langle n n' \rangle$ 
denotes pairs of nearest-neighbor sites.  We use the notation 
$K = \beta J$,  $h= \beta H$, 
\beq
a = z^{-1} = e^{K}
\label{a}
\eeq
and 
\beq
x = \frac{e^K-1}{\sqrt{q}}
\label{x}
\eeq
and denote the (reduced) free energy per site as 
$f = -\beta F = \lim_{N_s \to \infty} N_s^{-1} \ln Z$, where
$N_s$ denotes the number of sites in the lattice.  
Here we consider the square (sq) lattice.  There are actually $q$ types of
external fields which one may define, favoring the respective values 
$\sigma_n=0.,,,q-1$; it suffices for our purposes to include only one.  
The order parameter is defined to be
\beq
m = \frac{qM-1}{q-1}
\label{m}
\eeq
where $M = \langle \sigma \rangle = 
\lim_{h \to 0} \partial f/\partial h$.  With this definition, $m=0$ in the
$Z_q$-symmetric, disordered phase, and $m=1$ in the limit of saturated
ferromagnetic (FM) long-range order.  We shall refer to $m$ as the 
magnetization.  Finally, the (reduced, initial) susceptibility is denoted as 
$\bar\chi = \beta^{-1}\chi = \partial m/\partial h|_{h=0}$. 
We consider the zero-field model, $H=0$, unless otherwise stated. 

   The general $q$-state Potts model on the square lattice obeys the duality
relation \cite{potts} 
\beq
(a-1)(a_d-1)=q
\label{dualrel}
\eeq
where $a_d \equiv {\cal D}(a)$ is the image under the duality map ${\cal D}$ of
$a$; 
\beq
{\cal D}(a) = 1 + \frac{q}{a-1}
\label{dualmap}
\eeq
with ${\cal D}^2=1$ as usual.  In terms of the variable $x$, this duality
relation takes the simple form
\beq
x_d = \frac{1}{x}
\label{xdual}
\eeq
The critical point at which a phase transition occurs between the
high-temperature, symmetric phase and the low-temperature, FM phase, is given
by the self-dual point $x_c=1$, i.e., $a_c = 1+\sqrt{q}$. One may observe that
eq. (\ref{dualrel}) also has a second self-dual solution at a 
complex-temperature point, 
\beq
a = 1 - \sqrt{q} \ \equiv \ a_{\ell}
\label{aell}
\eeq
Exact expressions for the free and internal energy, latent heat, and 
magnetization have been  given by Baxter \cite{baxter73,baxter82} on the 
critical self-dual curve 
\beq
(a-1)^2=q
\label{fmcurve}
\eeq
Since the latent heat is zero at $a_c$ for $q \le 4$ \cite{baxter73}, the
corresponding transition between the symmetric and FM phases is continuous. 
The curve (\ref{fmcurve}) also includes the complex-temperature singular 
point at $a_\ell$ in eq. (\ref{aell}). 
We note that by the same reasoning, the phase boundary associated with the
complex-temperature point $a_\ell$ is also continuous.  Exact results have 
also been given by Baxter for the critical manifold \cite{baxter82,bc}
\beq
(a+1)^2 = 4-q 
\label{afmcurve}
\eeq
For $q=3$, eq. (\ref{afmcurve}) has two solutions: the AFM critical point at 
$a=0$, i.e. $T=0$ with $J < 0$, and a complex-temperature point, 
\beq
a = -2 \ \equiv \ a_s
\label{as}
\eeq
We observe that these two points are mapped into each other under the duality
mapping ${\cal D}$: 
\beq
{\cal D}(a=0) = -2
\label{duality}
\eeq
For $q=4$, eq. (\ref{afmcurve}) has only the solution $a=-1$, implying that for
$q \ge 4$ the model has no AFM long-range order even at $T=0$.

  Consider an $L_x \times L_y$ planar square lattice
$G$ with $N_s \equiv N_0$ sites (0-cells), $N_b \equiv N_1$ bonds (1-cells)
and $N_p \equiv N_2$ plaqettes (2-cells). We shall discuss boundary conditions
in section 3.5 below. The dual of $G$, which we denote
$G_d={\cal D}(G)$, is defined by associating uniquely a $(2-p)$-cell of $G_d$ 
with each
$p$-cell of $G$, so that $(N_0)_G=(N_2)_{G_d}$, $(N_1)_G=(N_1)_{G_d}$, and
$(N_2)_G=(N_0)_{G_d}$.  Since the planar graph has no handles, $N_h=0$ and
hence Euler characteristic $\chi_E=2(1-N_h)=2$, it follows from the 
Euler relation $\sum_{j=0}^2 (-1)^j N_j = \chi_E$ that $N_0-N_1+N_2=2$.  The
duality relation connecting the partition function $Z_G$ on $G$ with that
on $G_d$ is \cite{wurev}
\beq
Z_G(x) = x^{N_1}q^{N_0-1-(1/2)N_1}Z_{G_d}(x_d)
\label{zduality}
\eeq
It follows from (\ref{zduality}) that in the thermodynamic limit, the
singularities of that the free energy at a point $a$ and its dual image $a_d$
are the same.  In particular, 
\beq
f_{sing}(q=3;a=-2) = f_{sing}(q=3;a=0)
\label{fq3rel}
\eeq

   In discussing the complex-temperature phase diagram, it is convenient to 
use the Boltzmann weight $z$ and its inverse $a$, and the related variable, 
$x$.  For $q=3$, the exact results discussed above show that the phase 
structure 
for physical temperature (i.e. $0 \le a \le \infty$) consists of (i) the 
disordered, $Z_3$-symmetric PM phase for $0 < a \le 1+\sqrt{3}$; (ii) the FM 
phase for $1+\sqrt{3} < a \le \infty$; and (iii) an AFM critical point at 
$a=0$.  For $q \ge 4$, the physical phase structure consists only of the PM
phase for $0 \le a \le 1+\sqrt{q}$ and the FM phase for $1+\sqrt{q} < a \le
\infty$.  One defines the complex-temperature extensions (CTE's) of the PM 
and FM phases
by analytically continuing away from the respective segments of the positive
real $a$ axis.  Two rigorous properties are the following. First, because the
model has a high-temperature series expansion with finite radius
of convergence, it follows that the CTE of the PM phase occupies a finite 
neighborhood surrounding the point $a=1$.  Second, 
it is easy to show that for sufficiently large $|a|$, one 
is necessarily in the (CTE of the) FM phase.  To see 
this, let $a = \rho_a e^{i\theta_a}$; then 
\beq
K = \ln a = \ln \rho_a + i( \theta_a + 2\pi n)
\label{klog}
\eeq
where $n$ denotes the Riemann sheet of the logarithm and may be taken equal to
zero here.  It is clear that for sufficiently large $|a|=\rho_a$, the angle
$\theta_a$ makes a negligible contribution to $K$, so that (given that $d=2$ is
above the lower critical dimensionality for the FM transition), the system will
be in the FM phase.  This fact can be seen equivalently as a
consequence of the fact that the model has a low-temperature expansion with a
finite radius of convergence, so that there is a finite neighborhood of the
origin in the complex $z$ plane where it is in the (CTE of the) FM phase.
Henceforth, we shall generally refer to the complex-temperature extension of
the FM phase simply as the FM phase and similarly with the PM phase.

\section{ $q=3$ Potts Model}

   We shall discuss our methods in detail for the $q=3$ square-lattice Potts
model and then proceed to the higher $q$ cases.  We begin with
analyses of the low-temperature series expansions.  The series for the
partition function, magnetization, and susceptibility have been calculated to
order $z^{47}$ \cite{beg}.  
As in our previous studies of complex-temperature singularities of various
spin models, we have used both dlog Pad\'e and differential approximants to
analyze the series.  We fit the specific heat $C$, the magnetization $m$, and
the (reduced) susceptibility $\bar\chi$ to the leading singular forms
applicable near a generic singular point $z_{sing}$: 
\beq
C \sim (1-z/z_{sing})^{-\alpha_{sing}'}
\label{csing}
\eeq
\beq
m \sim (1-z/z_{sing})^{\beta_{sing}}
\label{msing}
\eeq
and 
\beq
\bar\chi \sim (1-z/z_{sing})^{-\gamma_{sing}'}
\label{chi}
\eeq
As usual, the primes indicate that we are approaching this singularity from
within the interior of the FM phase.   
We find convincing evidence from our series analyses for singularities at
two complex-conjugate points, which we denote $z_e$ and $z_e^*$, at which the 
magnetization, susceptibility, and specific heat are divergent. 

\subsection{Magnetization} 

 In Table 1 we present some diagonal and near-diagonal dlog Pad\'e results 
for this singularity.  (We have, of course, also calculated approximants 
farther from the diagonal.) 
\begin{table}
\begin{center}
\begin{tabular}{|c|c|c|} \hline \hline  & & \\
$[N/D]$ & $z_e$ & $\beta_e$ \\
 & & \\
\hline \hline
$[13/12]$ & $-0.340504 +0.287457 i$ & $-0.1258$ \\ \hline
$[13/13]$ & $-0.340420 +0.287491 i$ & $-0.1256$ \\ \hline
$[14/13]$ & $-0.340382 +0.287392 i$ & $-0.1250$ \\ \hline
$[13/14]$ & $-0.340358 +0.287473 i$ & $-0.1252$ \\ \hline
$[14/14]$ & $-0.340441 +0.287408 i$ & $-0.1254$ \\ \hline
$[15/14]$ & $-0.340428 +0.287382 i$ & $-0.1252$ \\ \hline
$[14/15]$ & $-0.340427 +0.287365 i$ & $-0.1251$ \\ \hline
$[15/15]$ & $-0.340433 +0.287339 i$ & $-0.1251$ \\ \hline
$[16/15]$ & $-0.340436 +0.287386 i$ & $-0.1253$ \\ \hline
$[15/16]$ & $-0.340429 +0.287368 i$ & $-0.1252$ \\ \hline
$[16/16]$ & $-0.340512 +0.287332 i$ & $-0.1256$ \\ \hline
$[17/16]$ & $-0.340405 +0.287364 i$ & $-0.1250$ \\ \hline
$[16/17]$ & $-0.340412 +0.287339 i$ & $-0.1250$ \\ \hline
$[17/17]$ & $-0.340406 +0.287354 i$ & $-0.1250$ \\ \hline
$[18/17]$ & $-0.340416 +0.287357 i$ & $-0.1251$ \\ \hline
$[17/18]$ & $-0.340410 +0.287350 i$ & $-0.1250$ \\ \hline
$[18/18]$ & $-0.340406 +0.287356 i$ & $-0.1250$ \\ \hline
$[19/18]$ & $-0.340407 +0.287363 i$ & $-0.1250$ \\ \hline
$[18/19]$ & $-0.340376 +0.287413 i$ & $-0.1251$ \\ \hline
$[19/19]$ & $-0.340312 +0.287325 i$ & $-0.1244$ \\ \hline
$[20/19]$ & $-0.340260 +0.287344 i$ & $-0.1241$ \\ \hline
$[19/20]$ & $-0.340273 +0.287344 i$ & $-0.1242$ \\ \hline
$[20/20]$ & $-0.340305 +0.287229 i$ & $-0.1241$ \\ \hline
$[21/20]$ & $-0.340285 +0.287321 i$ & $-0.1242$ \\ \hline
$[20/21]$ & $-0.340292 +0.287319 i$ & $-0.1243$ \\ \hline
$[21/21]$ & $-0.340326 +0.287269 i$ & $-0.1243$ \\ \hline
$[22/21]$ & $-0.340711 +0.286857 i$ & $-0.1246$ \\ \hline
$[22/22]$ & $-0.340855 +0.286878 i$ & $-0.1247$ \\ \hline
$[23/22]$ & $-0.340688 +0.286888 i$ & $-0.1246$ \\ \hline
$[22/23]$ & $-0.340982 +0.287191 i$ & $-0.1260$ \\ \hline
$[23/23]$ & $-0.340806 +0.287311 i$ & $-0.1262$ \\ \hline
\hline
\end{tabular}
\end{center}
\caption{Values of $z_e$ and $\beta_e$ from dlog Pad\'{e} approximants to 
low-temperature series for $m$ for $q=3$.}
\label{table1}
\end{table}
Our analysis of the series for the order parameter indicates complex-conjugate
singularities at approximately $z_e, z_e^* = -0.34 \pm 0.29i$ (the location 
will be discussed further below) where $m$ 
diverges with the exponent 
\beq
\beta_e = -0.125(1)
\label{betae}
\eeq
A plausible inference is that the exact value of this exponent is 
\beq
\beta_e = -\frac{1}{8}
\label{betaexact}
\eeq
We comment that although such a divergence in the order parameter is forbidden
in usual physical phase transitions, it can and does occur at
complex-temperature singularities.  Indeed, in our previous work we have 
noted several instances where the magnetization diverges at 
CT singularities.  For example, exact results show that $M$ diverges 
(like $(1+3u)^{-1/8}$) at the CT point $u=-1/3$ in the (zero-field) spin 1/2 
Ising model on the triangular lattice and at 
$u=u_e=-(3-2\sqrt{2})$ in the Ising model with $\beta H = i\pi/2$ on the 
square lattice (like $(1-u/u_e)^{-1/8}$) \cite{only}.

   The appearance of the exponent (\ref{betae}) at this CT singularity in the
square-lattice Potts model is intriguing, since $-1/8$ is not a simple 
(negative) multiple of any of the physical magnetic exponents in the model. 
This contrasts with the above-mentioned examples from the 2D Ising model, 
where, as is clear from the exact solution, the divergent magnetic exponents 
at $u=-1/3$ on the triangular lattice and at $u=u_e$ on the square lattice for 
$h=i\pi/2$ are precisely minus the common value of $\beta=1/8$ at the physical
PM-FM critical point.  Specifically, for the PM-FM 
transition in the 2D $q=3$ Potts model, the 
thermal and magnetic exponents are $y_t=1/\nu=1/\nu'=6/5$ and 
$y_h=28/15$, whence $\alpha=\alpha'=2-d/y_t=1/3$, $\delta=(d/y_h-1)^{-1}=14$, 
$\beta=1/9$, $\gamma=\gamma'=13/9$, and $\eta=4/15$ \cite{expon}.  
One recalls that the exponent $1/8$ does occur in the set of the conformal 
weights for the $m=5$ conformal field theory relevant to the 2D $q=3$ Potts 
model, viz., $h_{1,2}=h_{4,4}=1/8$, where 
$h_{p,q} = [((m+1)p-mq)^2-1]/[4m(m+1)]$ for $p=1,...,m-1$, $q=1,...,p$, 
and where the central charge is given by $c = 1 - 6/[m(m+1)]$ and has the 
value $c=4/5$ for this case \cite{cft}. 
However, the relation of this to the appearance of the exponent $-1/8$ at the
complex-temperature singularities $z_e, z_e^*$ is obscure, for several
reasons.  First, as we discussed in Ref. \cite{chisq}, there are violations of
basic scaling relations at complex-temperature singularities, so that it is not
clear how to apply conformal field theory to such singularities (since CFT
implies, among other things, such scaling relations).  
Secondly, since the Hamiltonian is not real at complex-temperature 
singularities, it is not obvious why the unitary rational conformal series 
is relevant to such singularities.  

\subsection{Susceptibility} 

In Table 2 we present our corresponding results from the dlog Pad\'e analysis
of the low-temperature series for the susceptibility. 
\begin{table}
\begin{center}
\begin{tabular}{|c|c|c|} \hline \hline  & & \\
$[N/D]$ & $z_e$ & $\gamma_e'$ \\
 & & \\
\hline \hline
$[15/15]$ & $-0.337324 +0.290677 i$ & $1.123$ \\ \hline 
$[17/15]$ & $-0.339008 +0.289393 i$ & $1.190$ \\ \hline
$[15/16]$ & $-0.339062 +0.288573 i$ & $1.173$ \\ \hline
$[16/16]$ & $-0.338269 +0.289319 i$ & $1.149$ \\ \hline
$[17/16]$ & $-0.338281 +0.289318 i$ & $1.150$ \\ \hline
$[18/16]$ & $-0.338121 +0.289177 i$ & $1.138$ \\ \hline
$[15/17]$ & $-0.337814 +0.289264 i$ & $1.125$ \\ \hline
$[16/17]$ & $-0.338280 +0.289318 i$ & $1.150$ \\ \hline
$[17/17]$ & $-0.338270 +0.289319 i$ & $1.149$ \\ \hline
$[18/17]$ & $-0.338227 +0.289314 i$ & $1.147$ \\ \hline
$[19/17]$ & $-0.338392 +0.289613 i$ & $1.163$ \\ \hline
$[16/18]$ & $-0.338168 +0.289263 i$ & $1.143$ \\ \hline
$[17/18]$ & $-0.338226 +0.289312 i$ & $1.147$ \\ \hline
$[18/18]$ & $-0.338259 +0.289325 i$ & $1.149$ \\ \hline
$[19/18]$ & $-0.338336 +0.289377 i$ & $1.154$ \\ \hline
$[20/18]$ & $-0.338338 +0.289242 i$ & $1.151$ \\ \hline
$[17/19]$ & $-0.338077 +0.289273 i$ & $1.138$ \\ \hline
$[18/19]$ & $-0.338332 +0.289360 i$ & $1.153$ \\ \hline
$[19/19]$ & $-0.338311 +0.289333 i$ & $1.152$ \\ \hline
$[20/19]$ & $-0.338355 +0.289392 i$ & $1.155$ \\ \hline
$[21/19]$ & $-0.338537 +0.289589 i$ & $1.170$ \\ \hline
$[18/20]$ & $-0.338310 +0.289320 i$ & $1.151$ \\ \hline
$[19/20]$ & $-0.338341 +0.289364 i$ & $1.154$ \\ \hline
$[20/20]$ & $-0.338168 +0.289333 i$ & $1.145$ \\ \hline
$[21/20]$ & $-0.337704 +0.289629 i$ & $1.126$ \\ \hline
$[22/20]$ & $-0.337349 +0.290843 i$ & $1.111$ \\ \hline
$[19/21]$ & $-0.338488 +0.291209 i$ & $1.196$ \\ \hline
$[20/21]$ & $-0.337205 +0.289884 i$ & $1.102$ \\ \hline
$[21/21]$ & $-0.340368 +0.286150 i$ & $1.080$ \\ \hline
$[20/22]$ & $-0.336698 +0.291093 i$ & $1.064$ \\ \hline
\hline
\end{tabular}
\end{center}
\caption{Values of $z_e$ and $\gamma_e'$ from dlog Pad\'{e} approximants to 
low-temperature series for $\bar\chi$ for $q=3$.}
\label{table2}
\end{table}
We have also carried out a similar study with (first-order, unbiased)
differential approximants, which yields the same value, to within the
uncertainty.  We determine the value of the specific heat exponent at the 
singularities $z_e, z_e^*$ to be 
\beq
\gamma_e' = 1.14(6)
\label{gammae}
\eeq
where the uncertainty represents a theoretical estimate from the scatter of
values among different Pad\'e  and differential approximants. 

\subsection{Specific Heat}

   To study the complex-temperature singularities in the specific heat, we 
have again carried out analyses with both Pad\'e and differential 
approximants.  As an illustration, we 
show in Table 3 our results from the latter.  Our notation is 
the same as in our earlier works, e.g. Ref. \cite{chisq}; $[L/M_0;M_1]$ is the
differential approximant to the generic function $\phi(z)$ obtained as the 
solution to the ordinary differential equation 
$Q_0(z)\phi(z) + Q_1(z) (z d/dz)\phi(z) = R(z)$, where $Q_0$, $Q_1$, and 
$R$ are polynomials of order $M_0$, $M_1$, and $L$.  
A review of the methods is given in Ref. \cite{tonyg}. 
\begin{table}
\begin{center}
\begin{tabular}{|c|c|c|} \hline \hline  & & \\
$[L/M_0;M_1]$ & $z_e$ & $\alpha_e'$ \\
 & & \\
\hline \hline
$[8/16;14]$ & $-0.3407788 +0.2885827 i$ &  $1.072$ \\ \hline
$[8/16;15]$ & $-0.3407563 +0.2878427 i$ &  $1.052$ \\ \hline
$[8/16;16]$ & $-0.3395542 +0.2877949 i$ &  $0.997$ \\ \hline
$[8/16;17]$ & $-0.3399969 +0.2874121 i$ &  $0.986$ \\ \hline
$[8/17;15]$ & $-0.3393125 +0.2891874 i$ &  $0.960$ \\ \hline
$[8/17;16]$ & $-0.3391475 +0.2886514 i$ &  $0.967$ \\ \hline
$[10/15;13]$ & $-0.3405761 +0.2887910 i$ &  $1.049$ \\ \hline
$[10/15;14]$ & $-0.3405514 +0.2889317 i$ &  $1.044$ \\ \hline
$[10/15;15]$ & $-0.3400580 +0.2896685 i$ &  $0.987$ \\ \hline
$[10/15;16]$ & $-0.3392558 +0.2894922 i$ &  $0.935$ \\ \hline
$[10/16;14]$ & $-0.3396813 +0.2885613 i$ &  $0.989$ \\ \hline
$[10/16;15]$ & $-0.3397607 +0.2885435 i$ &  $0.994$ \\ \hline
$[12/14;12]$ & $-0.3401737 +0.2893914 i$ &  $1.001$ \\ \hline
$[12/14;13]$ & $-0.3405823 +0.2891851 i$ &  $1.049$ \\ \hline
$[12/14;14]$ & $-0.3406363 +0.2897800 i$ &  $1.059$ \\ \hline
$[12/14;15]$ & $-0.3395286 +0.2896329 i$ &  $0.934$ \\ \hline
$[12/15;13]$ & $-0.3407441 +0.2896263 i$ &  $1.071$ \\ \hline
$[12/15;14]$ & $-0.3404562 +0.2891536 i$ &  $1.039$ \\ \hline
$[14/13;11]$ & $-0.3399260 +0.2888016 i$ &  $0.989$ \\ \hline
$[14/13;12]$ & $-0.3399969 +0.2885408 i$ &  $1.001$ \\ \hline
$[14/13;13]$ & $-0.3395575 +0.2895497 i$ &  $0.919$ \\ \hline
$[14/13;14]$ & $-0.3395242 +0.2893949 i$ &  $0.915$ \\ \hline
$[14/14;12]$ & $-0.3405031 +0.2889864 i$ &  $1.030$ \\ \hline
$[14/14;13]$ & $-0.3395982 +0.2892116 i$ &  $0.942$ \\ \hline
$[16/12;10]$ & $-0.3394602 +0.2889824 i$ &  $0.955$ \\ \hline
$[16/12;11]$ & $-0.3391883 +0.2891054 i$ &  $0.929$ \\ \hline
$[16/12;12]$ & $-0.3394838 +0.2892739 i$ &  $0.939$ \\ \hline
$[16/12;13]$ & $-0.3393726 +0.2892381 i$ &  $0.934$ \\ \hline
$[16/13;11]$ & $-0.3394488 +0.2892846 i$ &  $0.937$ \\ \hline
$[16/13;12]$ & $-0.3394364 +0.2893045 i$ &  $0.933$ \\ \hline
$[18/11;9]$ & $-0.3404461 +0.2882994 i$ &  $1.032$ \\ \hline
$[18/11;10]$ & $-0.3398013 +0.2888885 i$ &  $0.979$ \\ \hline
$[18/11;11]$ & $-0.3395252 +0.2888272 i$ &  $0.962$ \\ \hline
$[18/11;12]$ & $-0.3396986 +0.2891254 i$ &  $0.955$ \\ \hline
$[18/12;10]$ & $-0.3394656 +0.2892506 i$ &  $0.939$ \\ \hline
$[18/12;11]$ & $-0.3394441 +0.2891136 i$ &  $0.945$ \\ \hline
\hline
\end{tabular}
\end{center}
\caption{Values of $z_e$ and $\alpha_e'$ from differential approximants to
low-temperature series for (reduced) specific heat, $C/(k_BK^2)$, for $q=3$.}
\label{table3}
\end{table}
We determine the specific heat exponent to be 
\beq
\alpha_e' = 1.0(1) 
\label{alphae}
\eeq

   As can be seen from Tables 1-3, the 
magnetization, susceptibility, and specific heat series give 
consistent values for the location of the singularities $z_e,z_e^*$. 
Combining these, we infer that 
\beq
z_e, z_e^* = -0.339(2) \pm 0.289(2)i
\label{ze}
\eeq
For our comparison with the plots of zeros of the partition function, it will
be convenient to re-express this in terms of the $a$ and $x$ variables.  We
list the results in Table 4 (together with positions of the corresponding 
singularity for higher values of $q$, to be discussed later). 

\begin{table}
\begin{center}
\begin{tabular}{|c|c|c|c|} \hline \hline  & & & \\
$q$ & $z_e,z_e^*$ & $a_e,a_e^*$ & $x_e,x_e^*$ \\
 & & & \\
\hline \hline
3 & $-0.339(2) \pm 0.289(2)i$ & $-1.71(1) \mp 1.46(1)i$ & 
$-1.56(1) \mp 0.841(3)i$ \\ \hline
4 & $-0.288(2) \pm 0.270(2)i$ & $-1.85(1) \mp 1.73(1)$ & 
$-1.42(1) \mp 0.866(4)i$ \\ \hline
5 & $-0.251(2) \pm 0.251(2)i$ & $-1.99(1) \mp 1.99(1)i$ &
$-1.34(1) \mp 0.891(4)i$ \\ \hline
\hline
\end{tabular}
\end{center}
\caption{Values of $z_e$ and, correspondingly, $a_e$ and $x_e$ from analyses of
low-temperature series for magnetization, susceptibility, and specific heat. 
See text for further details.}
\label{table4}
\end{table}

   In passing, we observe that although we have found violations of scaling 
relations such as $\alpha+2\beta+\gamma=2$ and $\alpha'+2\beta+\gamma'=2$ in
our previous work at various complex-temperature singularities (e.g., Refs. 
\cite{chisq,chith}), in the present case, we obtain 
$\alpha_e'+2\beta_e+\gamma_e'=1.9 \pm 0.1$ so that, to within the 
uncertainties, this exponent relation is satisfied. 

\subsection{Singularities at Dual Images of $a_e,a_e^*$}

   A rigorous consequence of the duality of the model is that the free energy
also is singular, with the same singularity, at the points which are the 
dual images of the $a_e$ and $a_e^*$, namely, for the central values, 
${\cal D}(a_e), \ {\cal D}(a_e^*) = 0.141 \pm 0.462i$
or equivalently, ${\cal D}(x_e), {\cal D}(x_e^*) = -0.496 \pm 0.267$.  
These points lie in the (CTE)PM phase.  Note that since $|x_c|=1$ while $|x_e|
= 0.56$, the singularities at $x_e, x_e^*$ lie closer to the origin in the 
$x$ plane than the physical critical point. 

\subsection{Connection of $a_e, a_e^*$ Singularities with CT Phase Boundary 
${\cal B}$ }

   We would also like to relate these complex-conjugate CT singularities at
$z_e, z_e^*$, or equivalently, in the complex $a$ plane, at $a_e, a_e^*$, to 
the complex-temperature phase boundary ${\cal B}$. 
 From our previous studies on CT singularities \cite{ih,only,chith}, 
we formulated a conjecture that
whenever an arc or line segment of the phase boundary protrudes into, and ends
in, the FM phase, there is a divergence in $M$ at the endpoint of this arc.  
We proved that this divergence in $M$ implies also a divergence in
$\chi$ at the same endpoint \cite{chith}.  Besides the exact 
results alluded to above which exhibit this behavior, our calculations of
partition function zeros for the 2D higher-spin square-lattice Ising model 
\cite{hs}, in conjunction with the series analyses of Jensen, Guttmann, and 
Enting \cite{jge}, are consistent with the conjecture.  A natural extension 
of the conjecture is that the divergences which we have found in $m$ at these 
CT points in the square-lattice $q$-state Potts model indicate that these 
points are endpoints of arcs of points where $f$ is
non-analytic, i.e. arcs on the CT phase boundary ${\cal B}$.  

   In order to test this conjecture here, we have carried out new calculations
of complex-temperature zeros of the partition function for several different 
finite lattices, using transfer matrix methods, as in our earlier study of
partition function zeros for the 2D higher-spin Ising model \cite{hs}.  
At appropriate points we shall make comparison with previous computations of 
zeros for the square-lattice Potts model \cite{mm}-\cite{chw}. 
It is desirable to, and we shall,
restrict ourselves to lattices with duality-preserving boundary conditions 
(DBC's) \cite{wurev,mbook,chw}.  These guarantee that ${\cal D}(G)=G$
i.e. the dual of a (finite) lattice $G$ is (isomorphic to) the original 
lattice.  To discuss these, we recall eq. (\ref{zduality}) and the associated
definitions.  Note that 2D periodic, i.e. toroidal, boundary conditions do 
not preserve duality for a finite lattice. 
The boundary conditions are not uniquely specified by the condition that they
preserve duality.  One type, which we label as type 1, was discussed in
Ref. \cite{chw} (see their Fig. 1).  
We shall need a straightforward generalization of it to the
case of an $L_x \times L_y$  lattice with $L_x \ne L_y$, and we describe this
as follows.  Let the lattice be oriented with the $x$ and $y$ directions 
being horizontal and vertical, respectively.  Let all of the sites on the 
upper and right-hand edges, including the corners, connect along directions 
outward from the lattice to a common site adjoined to this lattice (so that 
the upper right corner connects to this adjoined point via bonds in both the 
$x$ and the $y$ directions), 
while all of the sites on the lower and left-hand edges, 
excluding the previously mentioned corners, have free
boundary conditions.  For the dual lattice, the special adjoined point may be
taken to lie to the lower left of the lattice.  In Ref. \cite{chw} it was noted
that from (unpublished) calculations of zeros with other DBC's, they obtained 
agreement with their conclusions from type 1 DBC's that for 
$Re(x) > 0$ the zeros lie on the circle $|x|=1$ \cite{chw}.  We have also used 
a set of DBC's different from type 1, which we denote as type 2 \cite{wunote}. 
For these, let the $L_x \times L_y$ lattice have periodic boundary conditions 
(PBC's) in the $x$ direction, so that the lattice may be pictured as a
cylinder with its axis oriented vertically.  
Now connect all of the sites on the upper edge of the
cylinder to a special point adjoined to the lattice, and let all of the sites 
on the lower edge of the cylinder have free boundary conditions. The dual
lattice is constructed in the usual way, assigning sites to each 2-cell of
the original graph, and adjoining the special point below the cylinder with the
stipulation that the points on the upper edge of the dual lattice have free
boundary conditions and the points on the lower edge connect to the adjoined
point. For both type 1 and type 2 DBC's, 
$N_0=L_xL_y+1=N_2$ and $N_1=2L_xL_y$ so that the prefactor
$q^{N_0-1-(1/2)N_1}=1$ in eq. (\ref{zduality}).  Note also that 
both type 1 and type 2 DBC's force some 
sites to have coordination number 3 rather than 4, and the adjoined point has
coordination number $L_x+L_y$ for type 1 DBC's and $L_x$ for type 2 DBC's.  
This is in contrast to periodic
boundary conditions, which violate duality but maintain equal coordination
number for all lattice sites.  

   A third type of DBC has recently been suggested to us by F. Y. Wu 
\cite{wunote}; it can be defined as follows, and will be denoted as DBC type 
3: consider 
an $L_x \times L_y$ lattice with $L_x=L+1$, $L_y=L$.  Let all of the sites on
the longer (horizontal) upper edge of the lattice be connected to a special
adjoined point, via $L+1$ bonds, and similarly, let all of the sites on the
lower edge of the lattice be connected by $L+1$ bonds to a second adjoined
point. Finally, connect the two adjoined points by a single bond and let the 
sites on the vertical edges of the lattice have free boundary conditions (in
the outward horizontal directions).  This lattice has $N_0=N_2=L(L+1)+2$ and
$N_1=2(L^2+L+1)=2(N_0-1)$ (so again, $q^{N_0-1-(1/2)N_1}=1$ in 
(\ref{zduality})).  Evidently, type 3 DBC's share greater similarity with 
type 1 than type 2 since in types 1 and 3 no subsets of edge sites have 
periodic boundary conditions, while in type 2 the sites on the vertical edges 
do have PBC's.  For a given value of $q$, the patterns of zeros which we have 
obtained with type 3 DBC's are indeed, similar to those with type 1.

\begin{figure}
\epsfxsize=3.5in
\epsffile{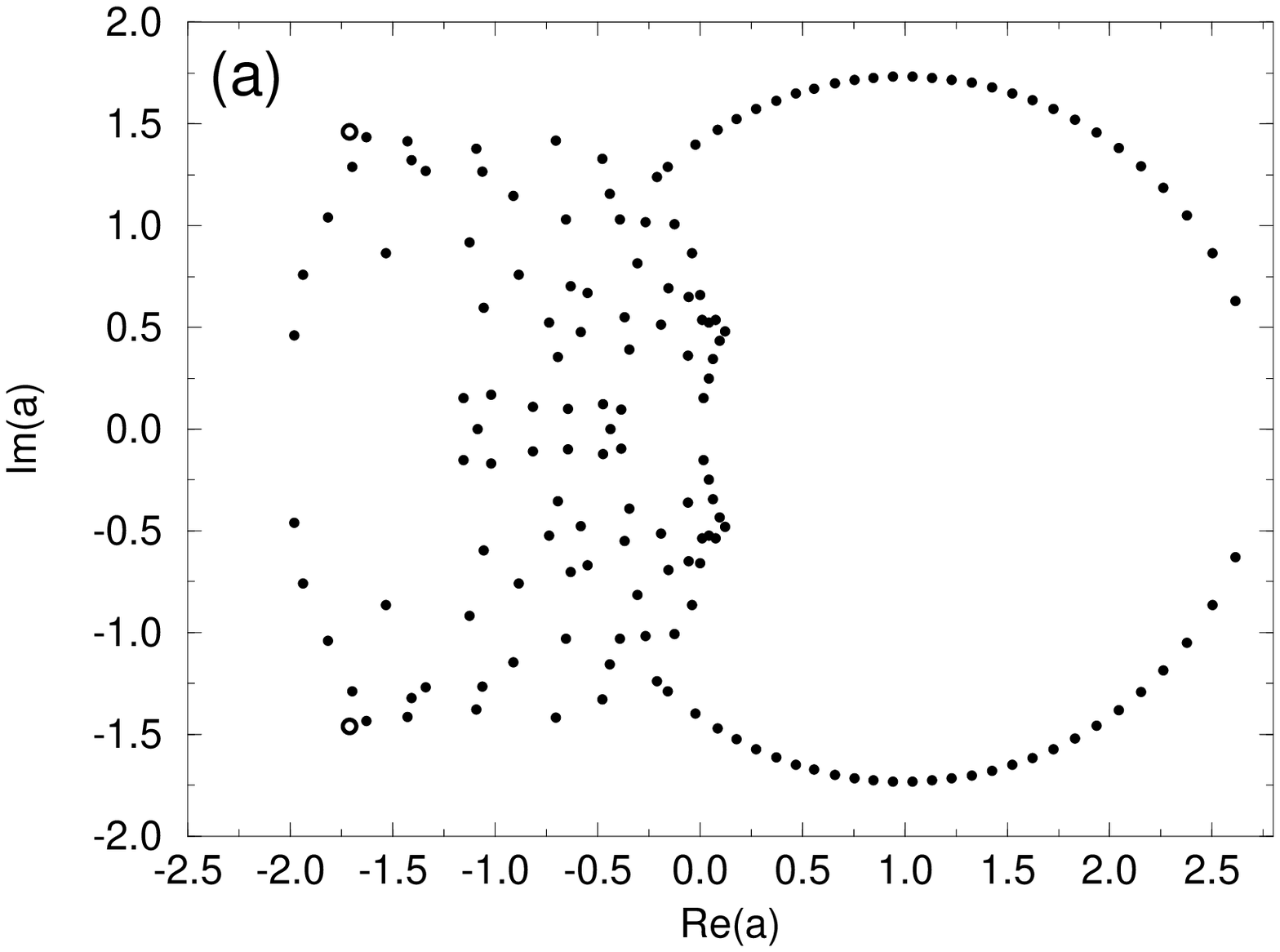}
\epsfxsize=3.5in
\epsffile{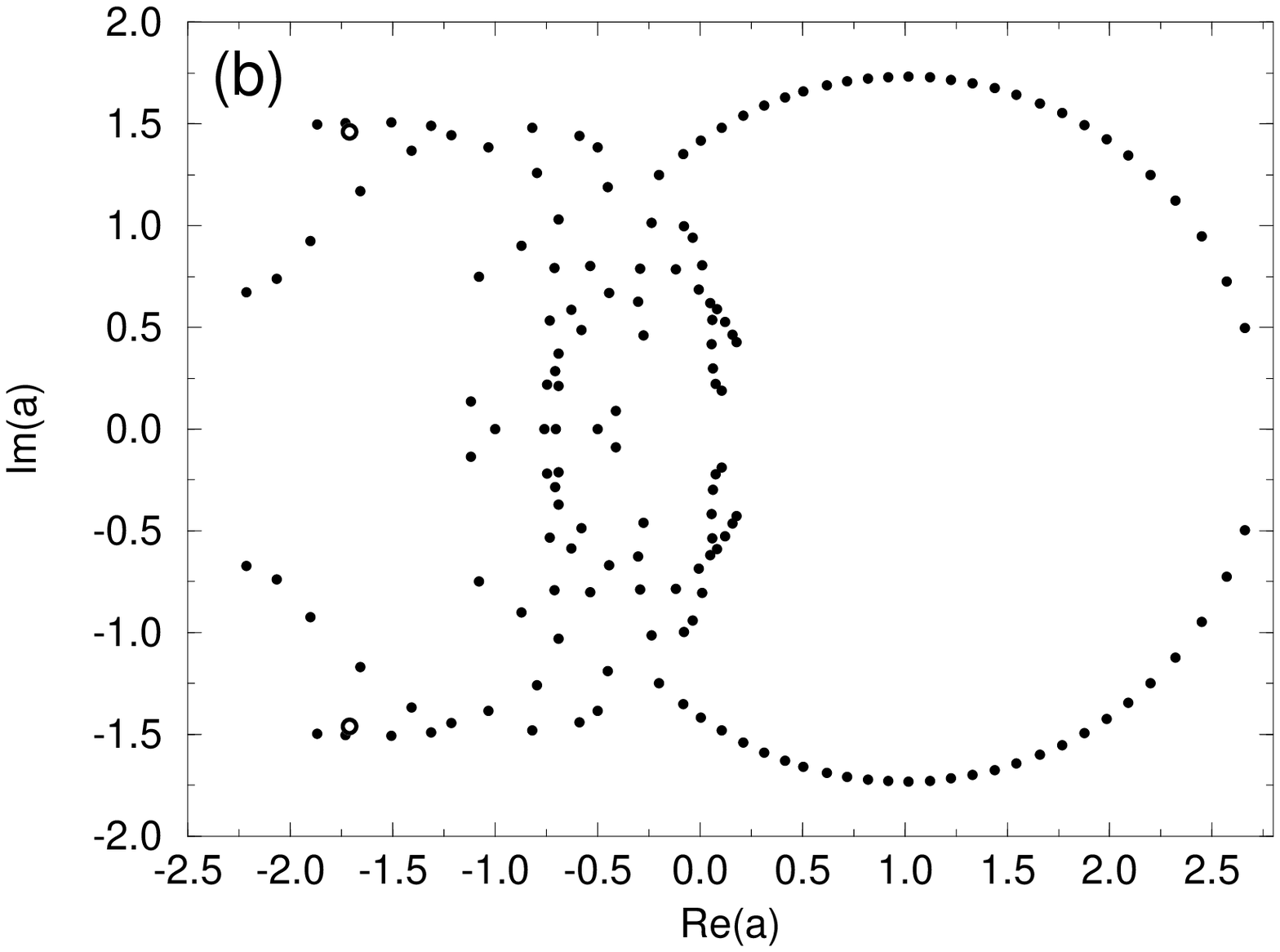}
\epsfxsize=3.5in
\epsffile{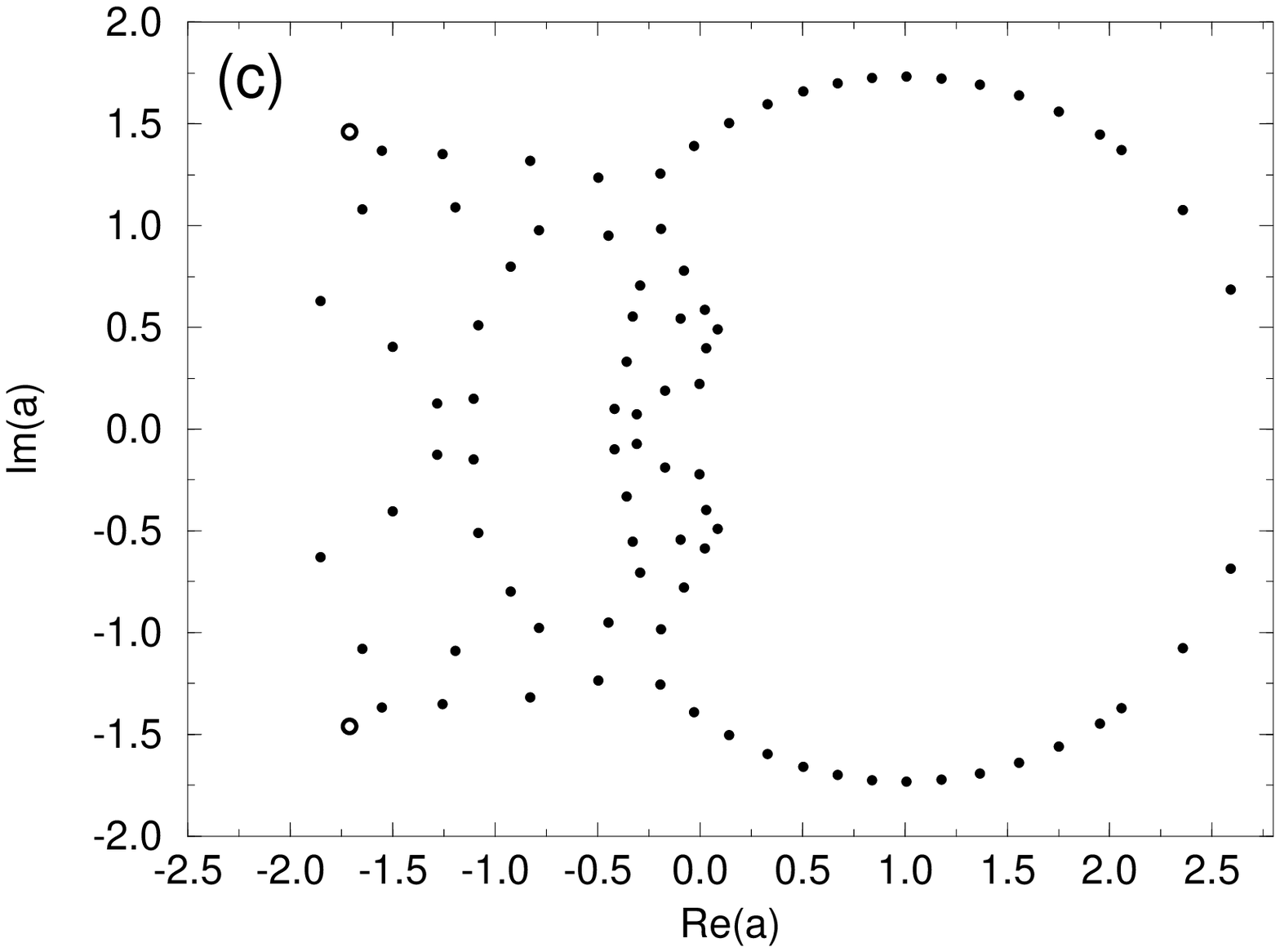}
\caption{Plot of zeros of $Z$, in the $a$ plane, for the square-lattice
$q=3$ Potts model
on an $8 \times 10$ lattice with duality-preserving boundary conditions 
(DBC's) of (a) type 1; (b) type 2; and (c) on a $7 \times 6$ lattice
with type 3 DBC's. The singularities $a_e,a_e^*$ are marked with $\circ$.}
\label{fig1}
\end{figure}

   In Figs. 1(a) and 1(b) we present calculations of complex-temperature 
zeros, in the complex $a$ plane, of the partition function for the $q=3$ 
Potts model on $8 \times 10$ lattices with duality preserving boundary 
conditions of types 1 and 2, respectively. The
positions of the singular points $a_e, a_e^*$ are marked with small circles on
both of these plots.  A comparison of these plots gives a quantitative 
measure of how the positions of the zeros can vary for different boundary
conditions, and specifically for those which maintain duality. This extends
previous published plots \cite{mbook,chw}, which showed that the pattern of 
zeros differs significantly when one uses duality-preserving, as opposed to
duality-violating, boundary conditions.  These comparisons demonstrated that 
once one specializes to duality-preserving BC's, the zeros in 
the $Re(a) > 0$ region lie nicely on the circle $|a-1|=\sqrt{q}$, whereas they
lie close to, but not exactly on, this circle for duality-violating BC's. 
Among previously published calculations with duality-preserving BC's, there 
is one, given as Fig. 11.1 in Ref. \cite{mbook}, on a lattice as large as 
the one that we use, and the pattern of zeros found there is very close to 
the pattern in our Fig. 1(b). In our figure, one can discern two outer-lying 
complex-conjugate arcs of zeros in the ``northwest'' and ``southwest'' 
quadrants, and the points $a_e,a_e^*$ lie near the ends of these arcs.  In
Fig. 1(a) there is more scatter among the zeros, 
but nevertheless, the points $a_e, a_e^*$ lie at the ends of 
subsets of zeros which can be associated with arcs. 
We have also performed analogous calculations of zeros on smaller
lattices with similar results.  For comparison, in Fig. 1(c) we show an 
exploratory calculation with type 3 DBC's.  Taking into account that the
lattice for Fig. 1(c) is somewhat smaller than that for Figs. 1(a,b), one
sees that the distribution of zeros is similar to that in Fig. 1(a) with
type 1 DBC's, especially
in the region near $a_e,a_e^*$ and their dual images.  From all of
these calculations, we thus conclude that the observed patterns of
zeros are consistent with the hypothesis that in the thermodynamic limit, the 
points $a_e, a_e^*$ are the ends of arcs contained in the complex-temperature
boundary ${\cal B}$ which protrude into the FM phase.  By duality, this is
equivalent to the statement that the singularities at ${\cal D}(a_e)$, 
${\cal D}(a_e^*)$ are the endpoints of arcs contined in ${\cal B}$ which
protrude into the PM phase. 

\subsection{Singularities at $a=0$ and $a=-2$} 

   The exact solution (\ref{afmcurve}) by Baxter shows that the point 
$a=0$, i.e. $T=0$ for $J < 0$, is the antiferromagnetic critical point.  By 
the duality property (\ref{zduality}), it follows that the free energy is 
also singular, with the same singularity, at the dual to this point, namely,
$a=-2 \equiv a_s$ given in eq. (\ref{as}). From renormalization group methods,
a mapping to a critical 6-vertex model, and studies of correlation functions,
it has been concluded that at the AFM critical point is an essential
singularity, with an essential zero
in the free energy and an exponential divergence in the correlation length
as $K \to -\infty$, i.e., $a \searrow 0$ \cite{ns}. Thus, if one assigns 
algebraic exponents $\alpha$ and $\nu$ for this AFM critical point, then
$\alpha = -\infty$, $\nu=\infty$.  By duality, the same singularity in the free
energy occurs at the dual point $a=-2$.  

   We have addressed two questions concerning these singularities at $a=0$
and $-2$: (i) how do they connect with the complex-temperature phase boundary
${\cal B}$, (ii) how well do the low-temperature series detect the CT singular 
point at $a=-2$?  For question (i), we first recall that 
the density $g$ of CT zeros along the curves comprising ${\cal B}$ in the
vicinity of a generic singular point $a_s$ behaves as \cite{abe}
\beq
g \sim |a-a_s|^{1-\alpha_s}
\label{gdensity}
\eeq
where $a_s$ denotes a singular point and $\alpha_s$ ($\alpha_s'$) denotes the
corresponding specific heat exponent for the approach to $a_s$ from within the
CTE PM (FM) phase.  Since $\alpha=-\infty$ at $a=0$ and hence, by duality,
$\alpha_s'=-\infty$ at $a_s=-2$, there is a strong reduction in the density of 
CT zeros as one approaches these respective points $a=0$ and $a=-2$ along
${\cal B}$.  This is consistent with what is observed with CT zeros 
calculated on finite lattices; one sees clear arcs of zeros and, e.g. for type
1 duality-preserving boundary conditions, these track toward the 
respective points $a=0$ and $a=-2$, ending some distance away from these
points.  (For type 2 DBC's, one also observes a slight curling tendency among 
the last few points on the curves.)   From this tracking of the zeros toward
$a=0$ (and hence, by duality, toward $a=-2$), we infer
that the AFM critical point $a=0$ lies on a portion of ${\cal B}$ 
which connects with the continuation of the curves lying on (either part or all
of) the unit circle $|a-1|=\sqrt{3}$ in such a way as to bound completely the 
complex-temperature extension of the PM phase.  This is 
reminiscent of another model which has no AFM long-range order at any
finite temperature, but an AFM critical point at $T=0$ \cite{stephenson}, 
namely, the (isotropic, spin 1/2) Ising model on the 
triangular lattice.  In that case, in terms of the variable $u=e^{-4K}$, the 
complex-temperature phase boundary ${\cal B}$ consists of the union of the 
circle $|u+1/3|=2/3$ and the semi-infinite line segment running along the 
negative real u axis from $-1/3$ to $-\infty$, or, equivalently, in the 
variable $u^{-1}$ analogous to $a$, it consists of the union of the circle 
$|u^{-1}-1|=2$ and the line segment $-3 \le u^{-1} \le 0$.  The AFM critical
point at $u^{-1}=0$ forms the right-hand end of this line segment and is
connected to the rest of ${\cal B}$ by it.  It is interesting to contrast 
this with situation in a model which is disordered and noncritical at
$K=-\infty$, i.e. at $u^{-1}=0$. 
An example is provided by the (isotropic, spin 1/2) Ising model on
the $3 \cdot 6 \cdot 3 \cdot 6$ (Kagom\'e) lattice \cite{not,suto}. Since the
point $u^{-1}=0$ is noncritical, it must be true that this point is not
connected to ${\cal B}$, and, indeed, one finds \cite{kag,cmo} that is lies in
the interior of the symmetric PM phase.  The same behavior is found for the $3
\cdot 12^2$ lattice \cite{cmo}.

   Returning to the $q=3$ Potts model, the property that the AFM critical point
at $a=0$ is connected to the rest of ${\cal B}$ as described above implies, by
duality, that the singular point at
$a=-2$ lies on the dual image of the above-mentioned curve, which therefore
encloses a complex-temperature phase and separates it completely from the (CTE
of the) FM phase.  Given the scatter of the zeros, it is not possible to make a
very reliable inference for where these curves intersect the continuation of
the circle $|a-1|=\sqrt{3}$ in the ``northwest'' and ``southwest'' quadrants of
the complex $a$ plane.  We do remark that the zeros are consistent with the 
possibility that these intersection points are $a=e^{\pm 2\pi i/3}$.  

   As regards the analysis of the low-temperature series, we find that these
series are able to locate the singular point at $a_s=-2$, i.e., $z_s=-1/2$, 
but not very
accurately.  Representative results for diagonal dlog Pad\'e approximants are
given in Table 5, again based on the series calculated to $O(z^{47})$ in 
Ref. \cite{beg}.  We do not show the results of the dlog Pad\'e approximants
$[N/D]$ with $N=D < 13$ because these did not locate the $z=-1/2$ singularity
with acceptable accuracy (e.g., the [12/12] and [13/13] approximants gave 
$-0.607830$ and $-0.572062$, respectively).  As is evident from Table 5, the
series yield values of the susceptibility exponent $\gamma_s' \sim 2$.  
Our previous work \cite{chisq} has shown that there are subtleties in
applying exponent relations like $\nu' d = 2 - \alpha'$ and 
$\gamma' = \nu'(2-\eta)$ at complex-temperature singularities.  However, it is
of interest to note that, since $\alpha_s'=-\infty$, if the hyperscaling
relation $\nu_s' d = 2 - \alpha_s'$ is valid, then $\nu_s'=\infty$ and hence,
assuming that the relation $\gamma_s'=\nu_s'(2-\eta_s)$ holds at $z_s$, it
would follow that $\gamma_s'=\infty$, i.e., $\chi$ would have an exponential 
divergence at this point.  Of course, the series analysis cannot yield
$\gamma_s'=\infty$, but it does produce a rather large value.  For 
reference, one may recall that in the case of the 2D O(2) model, while the 
Kosterlitz-Thouless theory implies that $\chi$ diverges exponentially as $T
\searrow T_c$, so that $\gamma=\infty$, the earlier series analyses gave a 
value of roughly $\gamma=3$ \cite{stanley}. 

\begin{table}
\begin{center}
\begin{tabular}{|c|c|c|} \hline \hline & & \\
$[N/D]$ & $z_s$ & $\gamma_s'$ \\
& & \\
\hline \hline
[14/14] & $-0.478716$ & 1.66  \\ \hline
[15/15] & $-0.490833$ & 2.31  \\ \hline 
[16/16] & $-0.487890$ & 2.12  \\ \hline 
[17/17] & $-0.487889$ & 2.12  \\ \hline
[18/18] & $-0.487906$ & 2.12  \\ \hline
[19/19] & $-0.487844$ & 2.11  \\ \hline
[20/20] & $-0.487995$ & 2.12  \\ \hline
[21/21] & $-0.485409$ & 1.99  \\ \hline
\hline
\end{tabular}
\end{center}
\caption{Values of $z_s$ and $\gamma_s'$ from diagonal dlog 
Pad\'{e} approximants to low-temperature series for $\bar\chi$ for $q=3$.} 
\label{table5}
\end{table}

\subsection{ Singularity at $a = 1 - 3^{1/2}$ }

  The Baxter solution (\ref{fmcurve}) shows that the free energy is also
singular at the self-dual complex-temperature point $a_\ell$ given in
eq. (\ref{aell}) for $q=3$, viz., $a = 1 - \sqrt{3}$. 
There is considerable scatter of zeros in the vicinity of this point (less for
Fig. 1(b) or Fig. 11.1 of Ref. \cite{mbook} than in Fig. 1(b)); however, the
zeros are consistent with the inference that this singular point lies on a
segment of the circle $|a-1|=\sqrt{3}$.  As one approaches the regions near the
intersection points discussed above, the scatter of zeros becomes too great to
draw a firm conclusion about this part of ${\cal B}$.  Nevertheless, we are 
able to infer 
that the point $a_\ell$ is completely separated from the FM phase by portions 
of the CT phase boundary ${\cal B}$.  To see this, assume the contrary,
i.e. that one can analytically continue from $z=0$ to $z_\ell=1/a_\ell$.
First, this would contradict the property that the singular point at $a=-2$
lies on a portion of ${\cal B}$ connecting it to the rest of ${\cal B}$ (the
dual image of the curve connecting the physical AFM critical point to the rest
of ${\cal B}$).  Second, if $a_\ell$ were not completely separated from the FM
phase, then one should be able to detect this singular point with the very 
long  low-temperature series available.  However, we found no evidence for a
singularity at this point from our analysis of these series.  The obvious
conclusion is that $a_\ell$ lies in a region beyond the applicability of these
series, i.e., beyond the border of the (CTE)FM phase.

\subsection{A Comment on $dim({\cal B})$ for $Re(a) < 0$}

 We comment here on another feature of the pattern of zeros. 
For the 2D spin 1/2 Ising model with isotropic 
spin-spin couplings, one knows from exact solutions that on most lattices, 
the zeros merge to form a 1-dimensional algebraic variety, i.e. the CT 
phase boundary ${\cal B}$.  Even for isotropic couplings, there is a 
heteropolygonal lattice, i.e., the $4 \cdot 8^2$ lattice, for which this is 
not the case; the locus of points where the free energy is non-analytic forms 
a 2-dimensional algebraic variety \cite{cmo}.  Moreover, for non-isotropic 
spin couplings, this is also true, even on the square lattice \cite{areas}. In
both cases it is easy to see why this is true (see section 6 of
Ref. \cite{cmo}).  The zeros of the 2D Ising model for higher spin values also
appear to approach curves as the lattice size gets large \cite{hs}.  For the 
2D $q$-state Potts model, the zeros in the $Re(a) > 0$ half-plane 
lie on a 1-dimensional curve, i.e., part of a circle
\cite{mm}-\cite{wuetal}.  In the $Re(a) < 0$ region, we are not
aware of any proof of this.  However, we can observe that in the known cases
with exact solutions, in the thermodynamic limit, the zeros either (i) form
curves, or (ii) areas, but not both curves and areas. 
Thus, given that the zeros for $Re(a) > 0$ do form a curve, one would have a
qualitatively new situation not previously encountered if some of the zeros in
the $Re(a) < 0$ did form areas.

\section{$q=4$ Potts Model}

    We have carried out exploratory analyses for higher-$q$ Potts models on 
the square lattice and have found evidence for singularities analogous to
$z_e,z_e^*$ in each of the cases studied.  We begin with the $q=4$ model.  
Like the $q=2,3$ cases, this model has a continuous, second-order PM-FM 
transition.  However, from the exact
solution by Baxter on the manifold (\ref{afmcurve}), it follows that the $q=4$
model does not have any AFM critical point.  Note that the complex-temperature
point $a=-1$ lies on the manifold (\ref{afmcurve}). 

    We have analyzed the low-temperature series for the magnetization,
susceptibility, and specific heat as before.  These yield a
consistent indication of singularities at the points listed in Table 4. 
(The values of $z_e$ from the $\bar\chi$ series exhibit somewhat greater 
scatter than those from the $m$ and $C$ series.) 
It is interesting that, to within the uncertainty, our numerical value for
$Im(a_e^*)$ can be fit by the exact expression $Im(a_e^*)=\sqrt{3}$ for 
$q=4$.  This and a similarly intriguing result which we find 
for $q=5$ suggest that there may be simple algebraic formulas for the
singularities $a_e,a_e^*$.  In view of the logarithmic confluent 
singularities which are known to occur at the physical PM-FM transition in 
this model \cite{conf}, it would be useful in future work to
carry out a more sophisticated analysis of the series including such
confluent singularities also at the complex-temperature singularities; however,
in the present exploratory work we have not done this.  From our study of the
low-temperature series for $m$, we find the exponent $\beta_e=-0.12(1)$, so
that, as before, the magnetization diverges at $z_e,z_e^*$.  This rigorously
implies \cite{chith} that $\chi$ also diverges at these points. 

\begin{figure}
\epsfxsize=3.5in
\epsffile{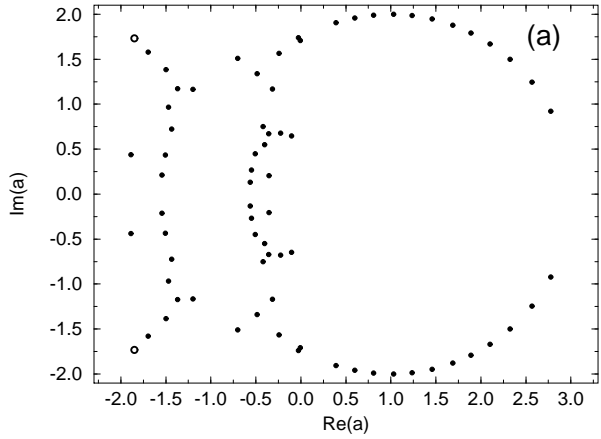}
\epsfxsize=3.5in
\epsffile{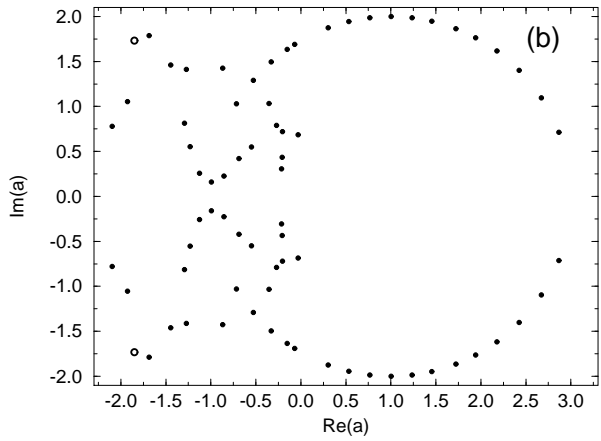}
\epsfxsize=3.5in
\epsffile{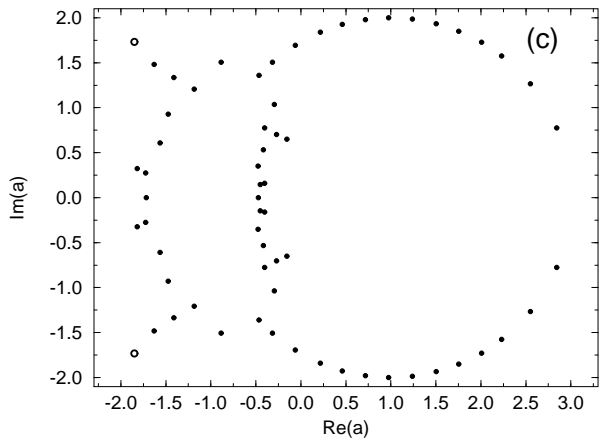}
\caption{Plot of zeros of $Z$, in the $a$ plane, for the square-lattice
$q=4$ Potts model on a $6 \times 6$ lattice with DBC of (a) type 1;
(b) type 2; and (c) on a $6 \times 5$ lattice with DBC's of type 3. 
Notation is as in Fig. 1.}
\label{fig2}
\end{figure}

   In order to see how these singularities connect to the complex-temperature
phase boundary ${\cal B}$ 
for the model, we have carried out a calculation of CT zeros 
of the partition function on various lattices with duality-preserving boundary
conditions.  In Figs. 2(a) and 2(b) we show calculations of these zeros, in 
the complex $a$ plane, for a $6 \times 6$ lattice with type 1 and type 2 
DBC's, respectively.  
Previously published calculations of zeros for this model include plots 
for strips ($L_x \times 32$ for $L_x=4$, 6, and 8; and $10 \times 16$)
\cite{m1} and a plot for a $4 \times 4$ lattice with type 2 DBC's.  We use 
symmetric lattices since in taking the thermodynamic limit on an 
$L_x \times L_y$ lattice, if $L_x/L_y$ deviates strongly from unity, the 
results can involve 1-dimensional artifacts.  For type 1 DBC's (Fig. 2(a)) 
one sees a clear indication of a complex-conjugate pair of arcs of zeros 
in the ``northwest'' and ``southwest'' quadrants, with
the singularities at $a_e, a_e^*$ forming the endpoints of these arcs.  
These arcs are not as clear with type 2 
DBC's ((Fig. 2(b)), but again, the points $a_e, a_e^*$ lie at 
the ends of subsets of zeros which can be associated with arcs.  A 
calculation with type 3 DBC's is included as Fig. 2(c).  The resultant pattern
is very similar to that in Fig. 2(a) with type 1 DBC's, to an even greater
extent than in the $q=3$ model. The patterns of zeros in Figs. 2(a)-2(c)
are all in good agreement with the
above-mentioned property that the model is not critical even at $a=0$, i.e.,
that it has no AFM critical point even at $T=0$.  For the plot in Fig. 2(b)
with type 2 DBC's, the zeros are consistent with the expectation from the 
Baxter solution ({\ref{afmcurve}) that the complex-temperature phase boundary 
${\cal B}$ passes through the point $a=-1$ (equivalently for this $q=4$ case, 
$x=-1$).  In contrast, for the plots in Figs. 2(a) and 2(c), with type 1 and
type 3 DBC's, respectively, there
are no zeros near to, or easily extrapolated toward, $a=-1$.  A clarification 
of the situation in the vicinity of this point merits further study on 
larger lattices. 

\section{Higher-$q$ Potts Model}

   We proceed to square-lattice Potts models with $q \ge 5$, for which the 
physical PM-FM phase transition is first-order.  Of course, a series analysis 
does not, in general, yield an accurate determination of the location of the
phase transition point for a first-order transition.  However, there is no
obvious reason why this should be a drawback for our study of the 
complex-temperature arc endpoint singularities $z_e,z_e^*$, since 
these yield a strong signal in the form of a divergent magnetization. 
In Table 6 we list the results for $z_e$ and corresponding
exponent $\beta_e$ from diagonal dlog Pad\'e approximants to the
low-temperature series for $m$.  We obtain similar results for $z_e$ from the
susceptibility and specific heat series. 
\begin{table}
\begin{center}
\begin{tabular}{|c|c|c|} \hline \hline  & & \\
$[N/D]$ & $z_e$ & $\beta_e$ \\
 & & \\
\hline \hline
$[10/10]$ & $-0.248723 +0.250655 i$ & $-0.1023$ \\ \hline
$[11/10]$ & $-0.248197 +0.250780 i$ & $-0.1004$ \\ \hline
$[10/11]$ & $-0.248617 +0.250694 i$ & $-0.1019$ \\ \hline
$[11/11]$ & $-0.249536 +0.249950 i$ & $-0.1034$ \\ \hline
$[12/11]$ & $-0.251131 +0.250021 i$ & $-0.1105$ \\ \hline
$[11/12]$ & $-0.251182 +0.250694 i$ & $-0.1142$ \\ \hline
$[12/12]$ & $-0.251032 +0.250976 i$ & $-0.1146$ \\ \hline
$[13/12]$ & $-0.250982 +0.250979 i$ & $-0.1143$ \\ \hline
$[12/13]$ & $-0.250978 +0.250974 i$ & $-0.1142$ \\ \hline
$[13/13]$ & $-0.251028 +0.250979 i$ & $-0.1146$ \\ \hline
$[14/13]$ & $-0.251528 +0.251159 i$ & $-0.1190$ \\ \hline
$[13/14]$ & $-0.251281 +0.250904 i$ & $-0.1157$ \\ \hline
$[14/14]$ & $-0.251527 +0.251148 i$ & $-0.1189$ \\ \hline
$[15/14]$ & $-0.251527 +0.251158 i$ & $-0.1190$ \\ \hline
$[14/15]$ & $-0.251475 +0.251197 i$ & $-0.1188$ \\ \hline
$[15/15]$ & $-0.251524 +0.251171 i$ & $-0.1190$ \\ \hline
$[16/15]$ & $-0.251529 +0.251161 i$ & $-0.1190$ \\ \hline
$[15/16]$ & $-0.251667 +0.251138 i$ & $-0.1199$ \\ \hline
$[16/16]$ & $-0.251668 +0.251140 i$ & $-0.1199$ \\ \hline
$[17/16]$ & $-0.251631 +0.251079 i$ & $-0.1191$ \\ \hline
$[16/17]$ & $-0.251667 +0.251138 i$ & $-0.1199$ \\ \hline
$[17/17]$ & $-0.251710 +0.251167 i$ & $-0.1205$ \\ \hline
$[18/17]$ & $-0.251396 +0.251142 i$ & $-0.1181$ \\ \hline
$[17/18]$ & $-0.251644 +0.251144 i$ & $-0.1198$ \\ \hline
$[18/18]$ & $-0.252009 +0.251663 i$ & $-0.1282$ \\ \hline
$[19/18]$ & $-0.251915 +0.251720 i$ & $-0.1267$ \\ \hline
$[18/19]$ & $-0.251910 +0.251717 i$ & $-0.1266$ \\ \hline
$[19/19]$ & $-0.251929 +0.251724 i$ & $-0.1270$ \\ \hline
\hline
\end{tabular}
\end{center}
\caption{Values of $z_e$ and $\beta_e$ from dlog Pad\'{e} approximants to 
low-temperature series for $m$, for $q=5$.}
\label{table6}
\end{table}
 From these diagonal (and near-diagonal) dlog Pad\'e approximants, we 
we obtain the value for $z_e$ listed in Table 4.  This is an intriguing result,
since, to within the uncertainty, our determination is consistent with the 
following exact analytic formula:
\beq
a_e,a_e^* = 2(-1 \mp i) \ \qquad {\rm for} \ \ q=5
\label{aeq5}
\eeq
Of course, in the absence of an exact solution of the model, we cannot 
exclude the possibility that this agreement is fortuitous, but it motivates 
one to think further about simple analytic expressions for the
location of the arc endpoints which we have discovered. 
 From our series analysis, we find that the magnetization diverges at these
points with exponent $\beta_e=-0.11(1)$.  As noted above, this implies
\cite{chith} that $\bar\chi$ must also diverge at these points, and our 
analysis of the low-temperature series for $\bar\chi$ yields the exponent
$\gamma_e'=1.2(1)$.

    We find the same generic features for all of the $q$ values
that we have analyzed, viz., complex-conjugate singularities at points $z_e,
z_e^*$.  These thus appear to be a general feature of 
the $q$-state Potts model on the square lattice for $q \ge 3$. 
Besides the case $q=5$, we have made exploratory studies of the
low-temperature expansions for the cases $q=6$, 7, and 8.  We find 
$z_e,z_e^*=-0.23 \pm 0.24i$, $-0.21 \pm 23i$, and $z_e=-0.20 \pm 22i$ for
$q=6,7$, and 8, respectively.  Thus, the magnitude $|z_e|$ ($|a_e|$) 
decreases (increases) as $q$ increases.  Note that if, in the $x$ plane, the
complex-conjugate arcs retract toward their respective points of origin and
finally disappear in the $q \to \infty$ limit, as is required in order for the
complex-temperature phase boundary ${\cal B}$ to reduce to the unit circle
$|x|=1$ in this limit, it is necessary that for large $q$, $|Re(a_e)|$ and 
$|Im(a_e)|$ grow like $\sqrt{q}$.  We find that $\beta_e$ increases from 
$\sim -0.11$ for $q=6$ to $\sim -0.10$ for $q=8$. 

\begin{figure}
\epsfxsize=3.5in
\epsffile{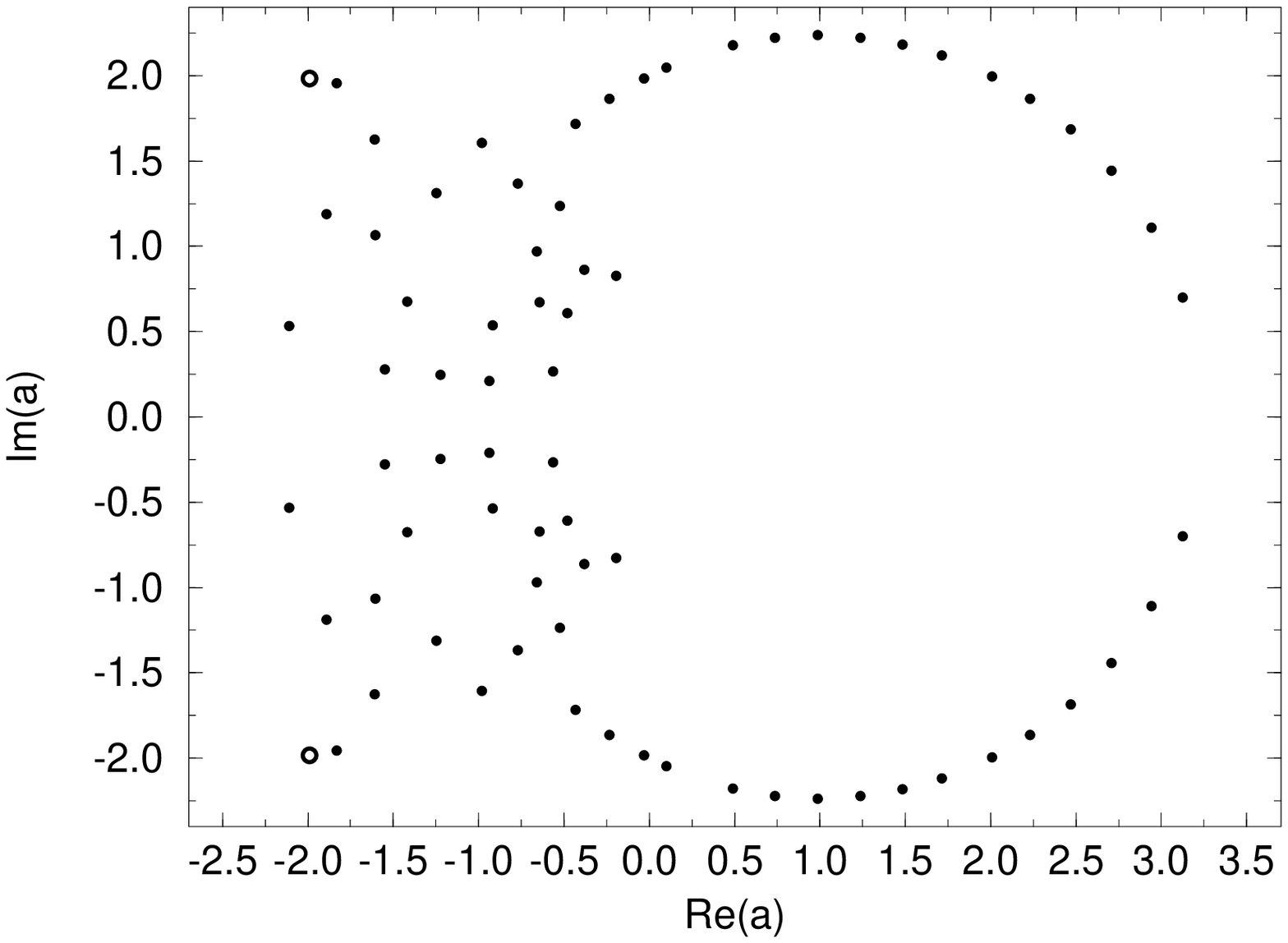}
\caption{Plot of zeros of $Z$, in the $a$ plane, for the square-lattice
$q=5$ Potts model on a $6 \times 6$ lattice with DBC of type 2. Notation is
as in Fig. 1.}
\label{fig3}
\end{figure}

\begin{figure}
\epsfxsize=3.5in
\epsffile{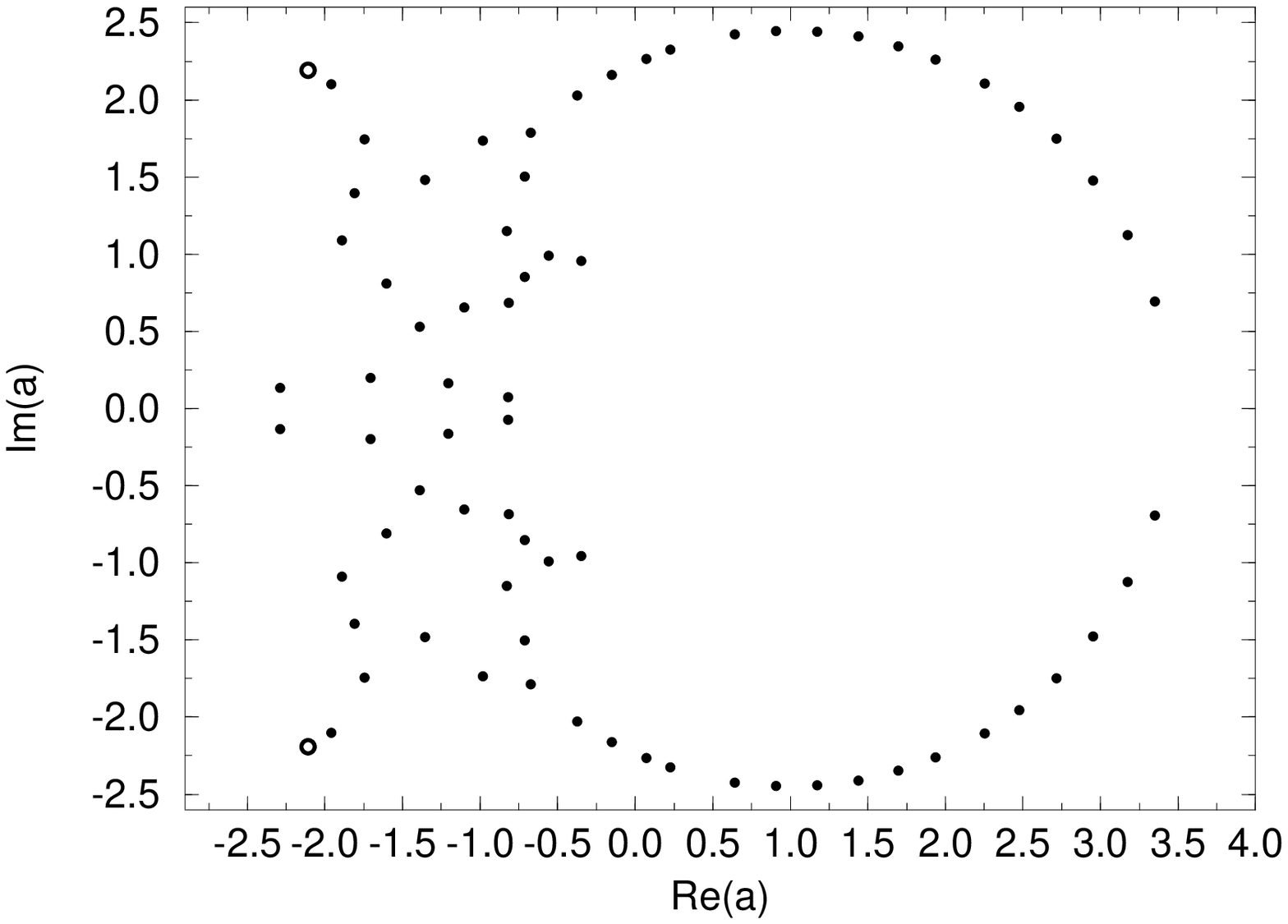}
\caption{Plot of zeros of $Z$, in the $a$ plane, for the square-lattice
$q=6$ Potts model on a $6 \times 6$ lattice with DBC of type 2. Notation is
as in Fig. 1.}
\label{fig4}
\end{figure}

In Figs. 3 and 4 we show our calculation of CT zeros of $Z$ for $q=5$ and $q=6$
on a $6 \times 6$ lattice with type 2 DBC's.  (We have checked that other DBC's
give results which also support our conclusions.)
Although there is considerable scatter of zeros in
the $Re(a) < 0$ region, the pattern is again consistent with the conclusion
that the singularities at $a_e,a_e^*$ lie at the end of arcs of zeros, so that
in the thermodynamic limit these points are endpoints of arcs of singularities
which connect with the rest of the complex-temperature phase boundary ${\cal
B}$.

\section{Conclusions}

   In summary, we have used analyses of low-temperature series expansions to
study complex-temperature singularities in the square-lattice $q$-state Potts 
model.  We have found singularities at complex-conjugate pairs of points and,
by means of comparison with patterns of partition function zeros, have obtained
support for the inference that in the thermodynamic limit these are endpoints
of arcs lying on the complex-temperature phase boundary ${\cal B}$.  At these 
points, the magnetization diverges, in agreement with an earlier conjecture
which we had formulated.  This guarantees that the susceptibility also diverges
at these points, and our series analyses are in accord with this. 
Our work includes several intriguing findings, including the
likely exponent value $\beta_e=-1/8$ for $q=3$ and an inference of an 
exact formula (\ref{aeq5}) for the endpoint singularities for $q=5$. From the
duality of the model, it follows that these arcs protruding into the 
complex-temperature extension of the FM phase are accompanied by their dual 
images, i.e. arcs protruding into the CT extension of the PM phase.  Our new 
results further elucidate the complex-temperature phase diagrams of 
square-lattice Potts models. 

\vspace{4mm}

   We thank Prof. F. Y. Wu for the helpful private communication 
\cite{wunote}. This research was supported in part by the NSF grant 
PHY-93-09888.

\vfill
\eject

\begin{thebibliography}{99}
\bibitem{not}{An Archimedean heteropolygonal lattice is defined by the symbol
$p_1 ... p_n$, meaning that as one makes a circuit around a site on the 
lattice, one traverses the polygons $p_1$, $p_2$, ... $p_n$; see Ref. 
\cite{cmo}.}
\end{thebibliography}
\end{document}